# Colossal photostructural changes in chalcogenide glasses. Athermal photoinduced polymerization in $As_xS_{100-x}$ bulk glasses revealed by near-bandgap Raman scattering

F. Kyriazis[1,2] and S. N. Yannopoulos[1,]*

[1] *Foundation for Research and Technology Hellas-Institute of Chemical Engineering and High Temperature Chemical Processes (FORTH-ICE/HT), P.O. Box 1414, GR-26504, Rio-Patras, Greece*

[2] *Department of Chemistry, University of Patras, GR-26504, Rio-Patras, Greece*

Near-bandgap Raman scattering was used to induce and study photostructural changes in $As_xS_{100-x}$ bulk glasses ($5 \leq x_{As} \leq 40$) revealing a new photoinduced polymerization effect. Raman spectra were recorded also in off-resonant conditions allowing for a detailed comparison between the "equilibrium" glass structure and the metastable one induced by illumination. It is shown that in S-rich glasses structural changes involve the athermal scission of $S_8$ rings and their polymerization to $S_n$ chains. The fraction of bonds involved in this effect is surprisingly high, being one order of magnitude higher than the corresponding fractions reported up to now in photostructural studies in chalcogenide glasses.

---

* E-mail: sny@iceht.forth.gr

Photoinduced atomic structural (PiS) changes are the hallmark of chalcogenide glasses (ChGs).[1,2] Such effects are also observed in elemental chalcogens and in particular in non-crystalline Se.[1,2,4] Because changes in structure, either at the short- or medium-range order, are inevitably related to alterations in macroscopic properties a vivid interest in structural exists in view of current technological applications of amorphous chalcogenides.[1] PiS changes in the As-S binary system have mainly been looked for in thin films.[2] Most studies have been focused on the stoichiometric compound, $As_2S_3$.[3] In the majority of cases the fraction of bonds which participate in PiS changes has been found to be of the order of ~2 % (Refs. 2, 5) while, in a Raman study,[3] an approximate analysis revealed a fraction of ~6% in $As_2S_3$. So far, PiS changes in stoichiometric binary glasses have been interpreted by the photolytic reaction |As-Ch|↔|As-As|+|Ch-Ch| (Ch: S, Se) where hetero-type bonds transform to homo-type ones after illumination, while the converse takes place by annealing. The rather small fraction of bonds involved in PiS changes implies a mild change of a material's property related to the population of the new "species" formed after illumination and thus entails a modest response of the material to an external stimulus. Because in most cases such changes are advantageous for applications, glasses which exhibit massive photoinduced bond restructuring are of great interest.

PiS changes are facilitated when glass structure contains atomic arrangements that are readily responsive to the incident radiation. Responsiveness depends upon two main conditions: (a) the ability of light to "excite" a bond between atoms, and (b) the possibility that the photo-excited state has to relax, non-radiatively, in a metastable configuration with different atomic arrangement compared with the initial one. Near bandgap light favors condition (a). Glasses with a rich variety of local micro-environments with alike bonding features are good candidates to fulfill condition (b).

Perhaps, S-rich As-S glasses are, among binary ChGs, the glasses whose structure displays the richest variety in different species. This stems from the fact that both $S_8$ rings and $S_n$ chains are the main building blocks in the structure of S-rich glasses, together with the $AsS_3$ pyramids. The above fact, i.e. adequate populations in the two different S micro-environments is exploited in the present work to reveal massive PiS changes related to the photoinduced scission and polymerization of $S_8$ towards $S_n$ chains.

Heat-induced polymerization in elemental S is a well-known effect usually referred to as the λ-transition of liquid sulfur.[6] $S_8$ rings open and polymerize abruptly at $T_\lambda \approx 159$ °C forming long chains with impact in many physicochemical properties of the liquid. An accurate determination of the temperature dependent polymer content of sulfur under various conditions has been recently achieved using Raman scattering.[7]

Photoinduced polymerization (PiP) of liquid S has attracted considerably less attention. On the experimental side, transient absorption experiments in liquid S revealed that polymerization of $S_8$ molecules takes place by bandgap illumination with a pulsed laser.[8] The relaxation processes through which photo-polymerized molecules return back to equilibrium was also studied in detail.[8] *Ab initio* molecular dynamics simulations followed experimental observation providing evidence for the photoinduced bond scission of isolated $S_8$ rings and partial polymerization in the liquid phase.[9] Apart form these works on PiP in elemental liquid S, no other study is known about this effect in sulfide glasses. It is therefore the aim of this work to focus on the details of the PiP effect in S-rich binary As-S glasses.

Binary $As_xS_{100-x}$ glasses with concentrations $x_{As}$: 5, 10, 15, 20, 25, 30, 33, and 40 at. % were prepared from high purity (5N) elements in an inert atmosphere glove box. Homogenized melts were quenched in water; the obtained glasses, apart from the two richest in S glasses ($x_{As}$ = 5 and 10), were well annealed near the corresponding glass transition temperature. Raman spectra were recorded with two set-ups using off-resonant and near-bandgap laser excitation: (i) a Fourier Transform Raman spectrometer (FRA 106/S, Bruker) operating at 1064 nm, and (ii) a dispersive Raman spectrometer (UV-LabRam, Jobin Yvon) equipped with the near-UV 441.6 nm laser line. In the latter, a laser power density of 0.09 mW was loosely focused in order to avoid thermal effects. In all cases we used the backscattering geometry and the same spectral resolution (3 cm$^{-1}$). The reason for selecting these wavelengths for Raman studies becomes obvious by considering that the 441.6 nm laser energy (2.81 eV) is comparable with the optical bandgaps of the S-rich glasses and in particular is resonant with the band gap of the $As_{10}S_{90}$ glass (2.79 eV).[10] However, the 1064 nm laser line is off-resonant for all glasses studied in this work and thus is used to reveal the true "equilibrium" (non-photo-perturbed) structure.

Figure 1 illustrates the Raman spectra of the As-S glasses excited by two laser wavelengths. Spectra are presented in the reduced representation where frequency and laser wavelength effects have been eliminated (see Ref. 7 for details on reduction of Raman spectra). The vibrational band in the spectral range 280-420 cm$^{-1}$ comes mainly from As-S symmetric and anti-symmetric stretching vibrations4 of $AsS_3$ pyramidal units that form upon alloying S with As. The spectra exhibit a fine structure in the off-resonant (1064 nm) Raman spectra at low As content where the interconnection of these units is limited. This fine structure is gradually lost and the band becomes systematically broader in the case of near-bandgap (441.6 nm) Raman scattering. The origin of this effect lies on resonance and/or pre-resonance effects that $AsS_3$ pyramidal units exhibit upon being illuminated by near-bandgap light. This is reasonable by envisaging that the glass structure at various As contents ($x_{As} < 33$) far from the stoichiometry exhibits nanoscale phase separation with S-rich and $AsS_3$-rich units. Preferential resonance effect of the 441.6nm radiation with the latter gives rise to the observed changes in Fig. 1. Emphasis is given here in the high energy Raman band which originates form S-S symmetric bond stretching modes.

Analysis in this spectral region[7,11] has revealed three distinct nanoscale environments that sulfur atoms can participate. (i) The band at ~474 cm$^{-1}$ represents the symmetric stretching mode of the $S_8$ rings. (ii) The band at ~492 cm$^{-1}$ grows strongly at $x_{As} > 25$ and is related to disulfide bonds, i.e. S-S bonds (or two-membered "$S^2$ chains") in $S_2$As-S-S-$AsS_2$ units. (iii) The band at ~463 cm$^{-1}$ has been identified with the symmetric stretching vibrational mode of polymeric chains.[7] In view of the above assignments the effect of near-bandgap light on the structure of sulfur-rich nanodomains is evident from Fig. 1. Normalization of the 474 cm$^{-1}$ (S-S bond stretching vibrational mode in a $S_8$ ring) peak intensities has been performed, in all but the richest in As glasses, in order to facilitate the comparison and the relative change of the 463 cm$^{-1}$ mode associated with $S_n$ chains. For S-rich glasses (i.e. $x_{As} \leq 25$) there is a considerable increase of the 463 cm$^{-1}$ band which entails a corresponding increase of the population of $S_n$ chains at the expense of $S_8$ rings. For $x_{As} \geq 25$ the population of $S_8$ has considerably decreased due to the presence of $AsS_3$ pyramids and there is therefore little possibility for $S_8 \rightarrow S_n$ transformations.

It might be crucial here to preclude any possibility that the $S_n$ band enhancement comes from either preferential resonance effects or heat-induced polymerization due to absorption of the laser radiation. Resonance effects between the various S-S bands are not expected in view of the proximity of the electronic configurations of S atoms in these species. This issue has been discussed in detail elsewhere[7(a)] in relation to the similar S-S bond polarizabilities in the two environments.[7(a)] Heat-induced polymerization in the course of near-bandgap illumination can also be safely ruled out due to the low power density used to record the Raman spectra. Moreover, a series of Raman spectra at off-resonant conditions were recorded for $As_5S_{95}$ as a function of temperature up to 250 °C.[7(c)] A comparison of high temperature spectra and the ambient temperature, near-bandgap Raman spectrum of the $As_5S_{95}$ glass revealed that the PiP effect is comparable in magnitude with heat-induced polymerization at 220 °C. At this temperature the Raman spectrum is much more different than the near-bandgap spectrum which ascertains the athermal nature of PiP.

The mechanism of PiP has been discussed in the corresponding effect in elemental S.[8,9] The group VI atoms have one s lone-pair, one p lone-pair and two p-orbitals to make a chemical bond with surrounding atoms. Absorption of the incident (near-bandgap) light serves to promote electrons from the lone-pair state to the anti-bonding σ* molecular orbital. This transition destabilizes the ring structure; hence rings open to form diradicals and concatenate to form chain molecules. To present in a quantitative way the structural changes responsible for the PiP effect we show in Fig. 2 the evolution of the population of S atoms participating in rings vs. $x_{As}$. The relative population has been estimated by the area ratio of the $S_8$ band (integration of the 220 cm$^{-1}$ band) over the total symmetric S-S band (integration of the 420-520 cm$^{-1}$ region). The ring population decreases almost linearly with increasing $x_{As}$ in the off-resonant case, almost vanishing at $x_{As} \approx 33$. On the other hand, near-bandgap light further decreases the $S_8$ population albeit in a non-linear fashion in this case. The difference curve shows that maximum ring destruction occurs at $x_{As} \approx 10$. Considering the relation between the optical bands gaps of the glasses and the laser wavelength this is an expected result in view of the proximity between the latter with the optical band-gap of the $As_{10}S_{90}$ glass.

An estimation of the fraction of bonds that are involved in the PiP effect can be done as follows. The random network[12] model, which is a good approximation for the structure of the As-S binary system, predicts that S-S bonds constitute almost 71.5% of the total number of bonds, at $x_{As} \approx 10$. Fitting the composite S-S band [440-520 cm$^{-1}$] using the procedure we followed in Ref. 7, we estimate that ~43% of the S-S bonds are in $S_8$ rings, which amounts to almost 31% of the total bonds in the glass. Figure 2 reveals a 65% reduction of the S-S bonds in rings upon illumination; which implies that *almost 20% of the total number of bonds in the glass is involved in the photo-induced structural changes responsible for PiP*.

In conclusion, we have revealed that an enormous fraction of S-S bonds contributes to the athermal PiP effect, i.e. scission of $S_8$ rings and polymerization to $S_n$ chains. This fraction is surprisingly high being almost one order of magnitude higher than the corresponding bond fractions reported up to now in studies of PiS changes in ChGs.[1-3] The key factor in the present case is the existence of a bimodal population of S atoms which participate in rings and chains and more importantly the fact that these two structurally different configurations have comparable populations at concentrations $5 \leq x_{As} \leq 15$.

We acknowledge financial support of the "PENED-03 / ΕΔ-887" project which is co-funded: 75% of public financing from the European Union – European Social Fund and 25% of public financing from the Greek State, Ministry of Development – GSRT; the Hellenic Telecommunications Organization (OTE S.A.) is also thanked for support.

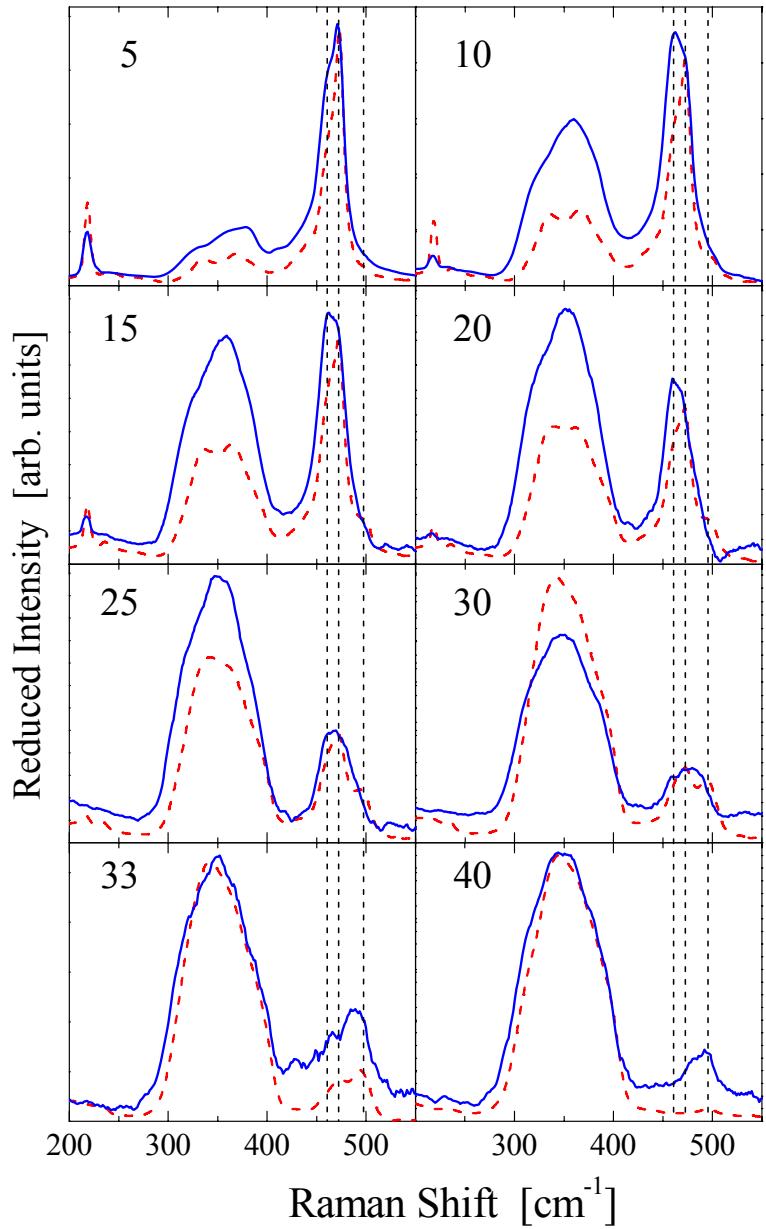

**Fig. 1:** Reduced Raman spectra of $As_xS_{100-x}$ glasses; numbers denote the As content in at. %. Blue solid lines: Raman spectra excited with the 441.6 nm laser wavelength; red dashed lines: FT-Raman spectra (1064 nm). Vertical dashed lines designate the peak positions of $S_n$, $S_8$ and $S^2$ species.

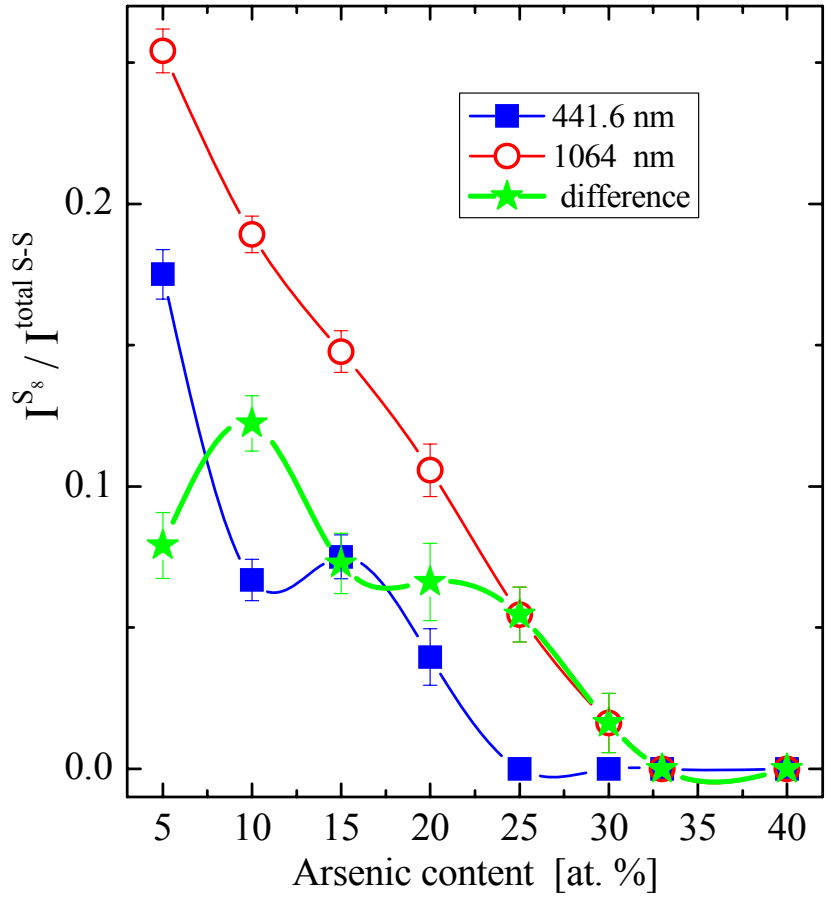

**Fig. 2:** Evolution of the intensity ratio $I^{S_8}/I^{S_{total}}$ (proportional to $S_8$ ring population) as a function of As content, open circles: off-resonant conditions; solid squares: near-bandgap conditions; solid stars: difference between these two cases.